# Graphene Spin Transistor

Sungjae Cho, Yung-Fu Chen[†] and Michael S. Fuhrer*

*Department of Physics and Center for Superconductivity Research, University of Maryland, College Park, Maryland 20742, USA*

*to whom correspondence should be addressed: mfuhrer@umd.edu

[†]present address: Department of Physics, 1110 West Green, Urbana, IL 61801-3080, USA

**Graphitic nanostructures, e.g. carbon nanotubes (CNT) and graphene, have been proposed as ideal materials for spin conduction[1-7]; they have long electronic mean free paths[8] and small spin-orbit coupling[9], hence are expected to have very long spin-scattering times. In addition, spin injection and detection in graphene opens new opportunities to study exotic electronic states such as the quantum Hall[10,11] and quantum spin Hall[9] states, and spin-polarized edge states[12] in graphene ribbons. Here we perform the first non-local four-probe experiments[13] on graphene contacted by ferromagnetic Permalloy electrodes. We observe sharp switching and often sign-reversal of the non-local resistance at the coercive field of the electrodes, indicating definitively the presence of a spin current between injector and detector. The non-local resistance changes magnitude and sign quasi-periodically with back-gate voltage, and Fabry-Pérot-like oscillations[6,14,15] are observed, consistent with**

**quantum-coherent transport. The non-local resistance signal can be observed up to at least $T = 300$ K.**

Figure 1 describes the spin-valve device (see Methods). Figure 1b shows the gate voltage ($V_g$) dependence of the resistivity ρ and conductivity σ. Similar to other single- and bi-layer graphene devices[10,16] $\sigma(V_g)$ shows a broad minimum around $4e^2/h$, where $e$ is the electronic charge and $h$ Planck's constant, increasing linearly with $V_g$ away from the minimum at $V_{cnp}$ (the charge neutrality point, CNP).

Figure 2a shows the four-probe non-local resistance $R_{nl} = V_{nl}/I$ (see Figures 1c,d) as a function of magnetic field $B$ at $V_g = +70$ V. $R_{nl}$ is positive at large $B$. As $B$ is swept to negative, $R_{nl}$ remains positive as $B$ crosses zero, then switches to a negative value at $B \approx -150$ G before returning positive at $B \approx -250$ G. Upon sweeping $B$ positive, switching occurs at $B = +150$ G and $+250$ G. This behavior is very similar to that observed in all-metal[13,17-19] and CNT[20] non-local spin-valves, particularly the sign change of $R_{nl}$ when the current and voltage circuits are separated[19,20]. Hence we identify these two magnetic fields as the coercive fields of F4 and F3 respectively. The switching behavior may then be explained as follows: at high $B$, F3 preferentially injects its majority spin which diffuses to F4 and is detected as an increase in the chemical potential of F4's majority spin (since the magnetizations of F3 and F4 are parallel) resulting in positive $R_{nl}$. When F3 and F4 are antiparallel, the voltage reverses, and $R_{nl}$ is negative.

Figure 2b,c,d shows the same measurement performed at different gate voltages and different electrode arrangements. In Figures 2b,d, the high-$B$ value is negative, and $R_{nl}$ switches to near zero (or slightly positive) for $B$ between the two coercive fields. The sign change is discussed further below. Figures 2e,f show the memory effect: two $R_{nl}$

states can be observed at $B = 0$, corresponding to the two possible magnetization states of the low-coercivity electrode.

First we discuss whether $R_{nl}$ arises due to charge current or spin current flowing between F3 and F4. Ideally, charge current would flow only between F3 and F2, eliminating contributions to the $R_{nl}$ from magnetoresistance of the ferromagnetic electrodes (anisotropic magnetoresistance), the channel, or the electrode-channel interface. However, because $R_{nl}$ is ~3 orders of magnitude smaller than the device resistance, it is possible that some charge current flows through a tortuous path from F3 to F4 and F5. We investigate this by measuring the gate voltage and temperature dependence of $R_{nl}$.

Figure 3a shows the gate voltage dependence of $R_{nl}$ in the parallel and antiparallel state, $R_{nl,p}$ and $R_{nl,ap}$, as well as their average value. Figure 3b shows the non-local spin-valve signal $\Delta R$. $R_{avg}$, $R_{nl,p}$ and $R_{nl,ap}$ all show a peak near the CNP (10 V $< V_g <$ 30 V), while $\Delta R$ is near zero in this region. Well outside this region ($V_g <$ -20 or $V_g >$ 40 V), $R_{nl,p}$ and $R_{nl,ap}$ have nearly equal magnitude and opposite sign ($R_{avg}$ is near zero) and $\Delta R$ is larger and shows quasi-periodic oscillations with $V_g$. The peak in $R_{avg}(V_g)$ near the CNP suggests that charge current *does* flow in the region between F3 and F4 for these gate voltages. However, $R_{avg}(V_g)$ is not simply proportional to $\rho(V_g)$ but rather drops to near zero at large $V_g$ while $\rho(V_g)$ remains finite. Thus the finite $R_{avg}(V_g)$ near the CNP is likely due to the inhomogenous nature of graphene near the CNP[21,22]; here percolating electron and hole regions may cause a tortuous current path.

Away from the CNP, $R_{avg}(V_g)$ drops to near zero, indicating small charge current. Yet $R_{nl,p}$ and $R_{nl,ap}$ remain finite, with near equal magnitude and opposite sign. This is as expected for a pure spin current flowing from F3 to F4, and cannot be explained by a magnetoresistive signal arising from any charge current between F4 and F5. The Hall effect is another possible source of $V_{nl}$, however, the Hall voltage would be expected to grow large and switch sign near the CNP, rather than showing a peak.

Figure 4 shows the temperature dependence of $R_{avg}$ and $\Delta R$ for $V_g = 0$. Here $R_{avg}$ is finite similar to Figure 3, but somewhat larger for this electrode configuration. The spin-valve signal $\Delta R$ is seen to drop with temperature approximately as $\Delta R \propto T^{-1}$, while $R_{avg}$ is much more weakly temperature dependent; again indicating a different origin for $\Delta R$ and $R_{avg}$. The inset shows a measurement at 300 K performed at higher current; the spin-valve signal can still be observed, confirming expectations of reduced spin scattering in graphene even to high temperature.

We now discuss the magnitude of the spin-valve signal $\Delta R$. For an Ohmically-contacted spin-valve device, the non-local signal may be estimated using Eqn. 22 of reference [17]; we estimate in this case the signal should be on order $10^{-5}$ Ω. However, we observe finite contact resistance of order 10 kΩ per electrode as estimated from the difference between two-probe and four-probe resistance measurements. In the limit of highly resistive contacts, we would expect the non-local resistance to be given by Eqn. 2 of reference [19], which, for long spin-scattering lengths, is on the order of the channel resistance (1-10 kΩ). Our intermediate contact resistance, finite spin-scattering length, and finite polarization of the electrodes will give a lower value of $\Delta R$, similar to the observation of $\Delta R \sim 20$ Ω for a channel resistance 10 kΩ and contact resistance of a few tens of kΩ in a CNT device[20].

We now discuss the origins of the quasi-periodic oscillations of the non-local spin-valve signal $\Delta R(V_g)$. Oscillation of the spin-valve signal with $V_g$ due to spin-orbit coupling has been proposed as the basis of a spin transistor[23]. However, the spin-orbit coupling in graphene is expected to be very small[9], and this effect should not be observable[24]. Oscillations and sign changes of the spin-valve signal have also been observed when the spin current flows through a resonant quantum state, either due to Coulomb blockade[25] or resonant tunneling[6]. It is evident from Figure 1b that the sample is not in the Coulomb blockade regime, however $\rho(V_g)$ shows quasi-periodic oscillations. We examined similar oscillations in another graphene sample in a two-probe geometry, (Figure 4c). In the color-scale plot of differential conductance vs. $V_g$ and drain voltage $V$, conductance maxima and minima occur along diagonal lines. In CNTs, such behavior is attributed to Fabry-Pérot interference of electronic states reflected from the electrodes[6,14], and similar behavior has been reported in graphene[15]. The round trip phase experienced by an electron traveling the CNT channel varies by $2\pi$ when the change in electron wavenumber is $\Delta k = \pi/L$. This results in a period in gate voltage $\Delta V_g = \frac{2}{L}\left(\frac{\pi e V_g{'}}{c_g}\right)^{1/2}$ and in drain voltage $\Delta V = \frac{hv_F}{eL}$. The slope of the lines is

$$\frac{\Delta V}{\Delta V_g} = \frac{2c_g}{e^2}\frac{1}{D(E)} = hv_F\left(\frac{c_g}{4\pi e^3 V_g{'}}\right)^{1/2}$$

, where $v_F = 1 \times 10^6$ m/s is the Fermi velocity, and $V_g{'} = |V_g - V_{cnp}|$. For this device, $L = 200$ nm, $V_{cnp} \approx -6$ V, giving $\Delta V = 20$ mV. For $V_g{'} = 10$-$15$ V, we find $\Delta V_g = 2.0$-$2.5$ V, and the slope varies from 0.013 to 0.010. The slopes match quite well the experimentally observed slopes (see Figure 3c). The most prominent minima in conductivity at $V = 0$ occur with spacing $\Delta V_g = 1.5$-$2.5$ V, however additional features are observed more closely spaced in $V$ and $V_g$ than expected from above. This is not surprising due to the two-dimensional nature of graphene: our analysis includes only the *k*-states perpendicular to the electrodes, undercounting the states involved in transport (the slope is independent of the path length *L*). The spin-valve sample also shows oscillations of $\sigma(V_g)$ in Figure 1b. For this sample $L = 450$ nm, and we would expect $\Delta V_g = 2.8$ V at $V_g' = 90$ V (i.e. $V_g = -70$ V) which agrees reasonably well with the observed spacing of dips $\Delta V_g \sim 6$ V at large negative $V_g$ in Figure 1b. The four-probe geometry is significantly more complicated than the two-probe analysis of Fabry-Pérot interference above, since there are multiple interfaces which could give rise to interference. Still it is reasonable that quantum interference effects are responsible for the oscillations in the four-probe resistivity (Figure 1b), and for the observed changes in magnitude and sign of the spin-valve signal with gate voltage (Figure 3b).

In conclusion, we have observed the non-local resistance arising from a spin current in graphene in a non-local four-probe measurement. The spin-valve signal varies with gate voltage in magnitude and sign due to interference arising from the quantum-coherent transport through graphene, which is also evidenced by Fabry-Pérot-like interference patterns observed in a similar sample, and oscillations in the four-probe resistivity of the spin-valve sample. The magnitude of the spin-valve signal is roughly inversely proportional to temperature, and is observable at room temperature. Injection and detection of pure spin currents in graphene opens possibilities to examine theoretically predicted new phenomena such as the spin Hall effect[9] and half-metallicity in graphene ribbons[12]. Because of the high current-carrying capability and long mean-free path at room temperature, graphene is also an excellent candidate for room-temperature spintronics applications.

**Acknowledgements**

This work has been supported by the U.S. ONR grant N000140610882, NSF grant CCF-06-34321, and the UMD-NSF-MRSEC grant DMR-05-20471.

**Methods**

Our graphene samples are obtained by mechanical exfoliation[26] on 300 nm $SiO_2$/Si substrates. We use optical microscopy to locate the graphene and verify single-layer thickness[27]; the optical contrast is compared with other samples fabricated on identical substrates which show the half-integer quantum Hall effect characteristic of graphene[10,11,21]. We estimate that all the samples discussed in this manuscript are at most two graphene layers thick. Ferromagnetic Permalloy electrodes are formed by electron-beam lithography (EBL) followed by thermal evaporation; a second EBL step establishes contact to the Permalloy via normal Cr/Au electrodes.

**Figure Captions**

**Figure 1 Graphene spin-valve device.** **a**, Optical micrograph of graphene on $SiO_2$/Si substrate. Scale bar is 10 microns, white box shows the graphene flake used in this study, which has similar contrast to other graphene samples for which half-integer quantum Hall effect was measured. **b**, Gate voltage ($V_g$) dependence of four-probe resistivity ρ (black, left scale) and conductivity σ (blue, right scale) at a temperature of 1.25 K. The field-effect mobility $\mu_{FE} = (1/c_g)|d\sigma/dV_g|$ is approximately 2500 cm$^2$/Vs, where $c_g = 1.15 \times 10^{-4}$ F/m$^2$ is the gate capacitance. In this local resistivity measurement, electrodes F4 and F5 were used as voltage probes, and the current contacts were F3 and F6. **c,d**, Schematics of device layout. **c**, Plan view. **d**, Side view, showing setup for non-local resistance measurement. Six ferromagnetic Permalloy electrodes F1-F6 were deposited on top of the graphene strip. To give different coercive fields, F1, F3, F5 have dimensions 0.4 μm x 15 μm, and F2, F4, F6 are 1.0 μm x 3 μm. Spaces between all the electrodes are 450nm.

**Figure 2 Non-local spin-valve effect in graphene.** **a,b**, Nonlocal resistance $R_{nl}$ (see **Figure 1d**) as a function of magnetic field measured at temperature $T$ = 20K with current $I$ = 100 nA. **a,b** F2, F3 as current leads; F4, F5 voltage leads. **c,d** F4, F5 as current leads; F2, F3 voltage leads. **a**, Gate voltage $V_g$ = +70 V. **b**, $V_g$ = -67 V. **c**, Gate voltage $V_g$ = -20 V. **d**, $V_g$ = -69 V. The non-local resistance switches sign upon sweeping magnetic field, which indicates that a spin current flows from electrodes F3 to F4 (see **Figure 1c,d**). The reversal of sign of the non-local resistance with gate voltage (**a** vs. **b** and **c** vs. **d**) is discussed in the text and in **Figure 3**. **e,f**, Memory effect measured at $T$ =

20 K, $I = 100$ nA, $V_g = 0$ V.  F1, F2 are current leads; F3, F4 are voltage leads.  In **a-f**, blue curves correspond to positive sweep direction of magnetic field; black curves, negative sweep direction.

**Figure 3 Gate-voltage dependence of spin-valve signal.  a**, Resistance as a function of gate voltage for electrodes with magnetizations parallel ($R_{nl,p}$), antiparallel ($R_{nl,ap}$), and their average $R_{avg} = (R_{nl,p} + R_{nl,ap})/2$.  **b**, The spin valve signal $\Delta R = R_{nl,p} - R_{nl,ap}$ as a function of gate voltage.  **a,b,** The same electrode configuration is used as for **Figure 2a,b**.  **c**, Color-scale plot of two-probe differential conductance as a function of gate voltage $V_g$ and drain voltage $V$ for a similar graphene sample contacted by Permalloy electrodes with a spacing 200 nm.  The blue and red dashed lines have slopes of $\pm 0.013$ and $\pm 0.010$ respectively.

**Figure 4.  Temperature dependence of spin-valve signal.**   The average resistance $R_{avg}$ (black circles) and spin valve signal $\Delta R$ (blue squares) as a function of temperature at $V_g = 0$ V and $I = 100$ nA. The temperature dependence of the spin-valve signal $\Delta R$ is much stronger than that of $R_{avg}$ (which likely arises from the charge-current resistance).  The solid line shows a power law with exponent -1.  Inset shows the non-local resistance $R_{nl}$ as a function of field at $T = 300$ K and $I = 3$ µA. Blue curve is positive sweep direction of magnetic field; black curve, negative sweep direction. The spin-valve effect is still observable.  The electrodes used for the spin-valve data in main panel and inset are the same as for **Figure 2e,f**.

Figure 1

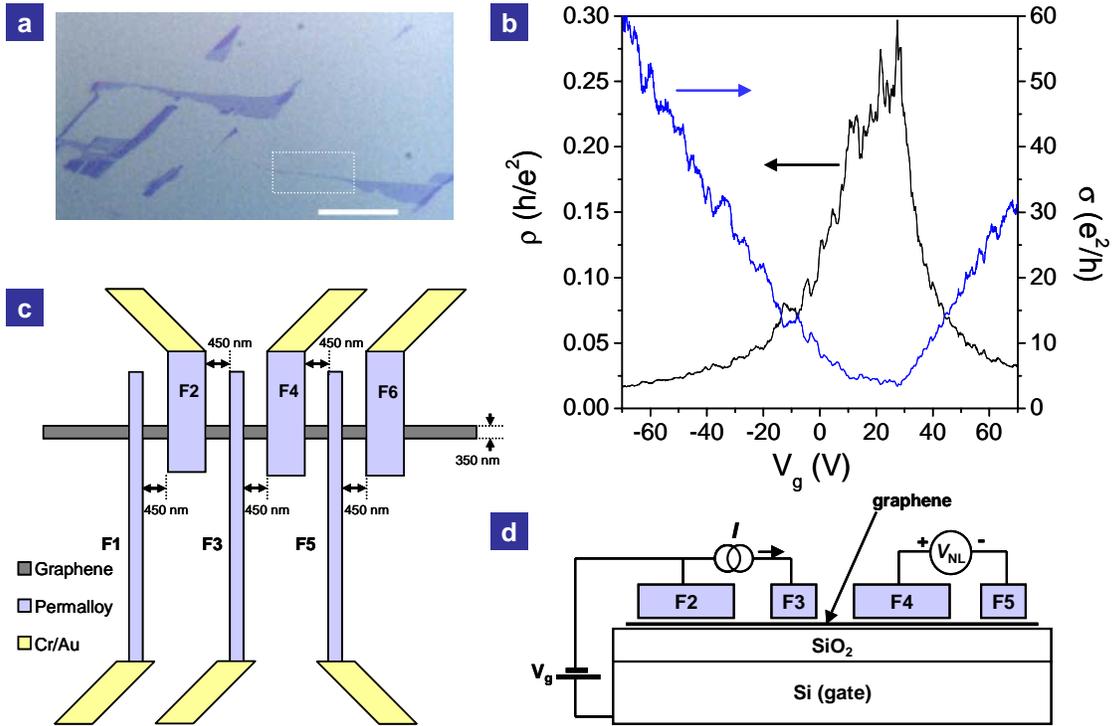

Figure 2

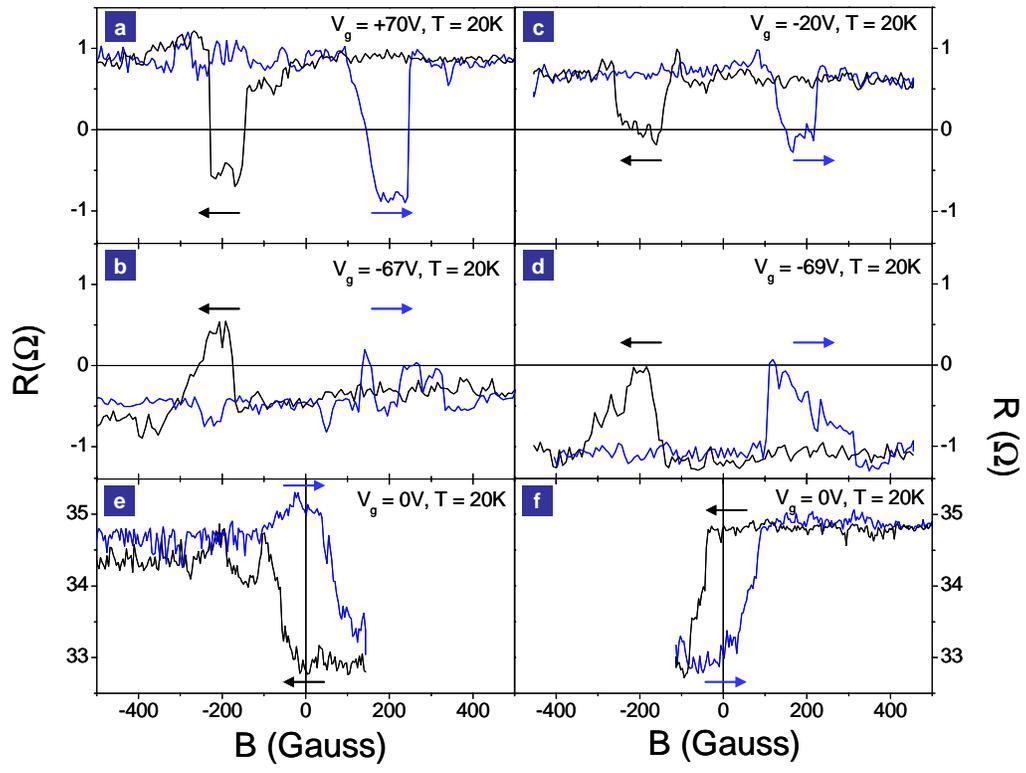

Figure 3

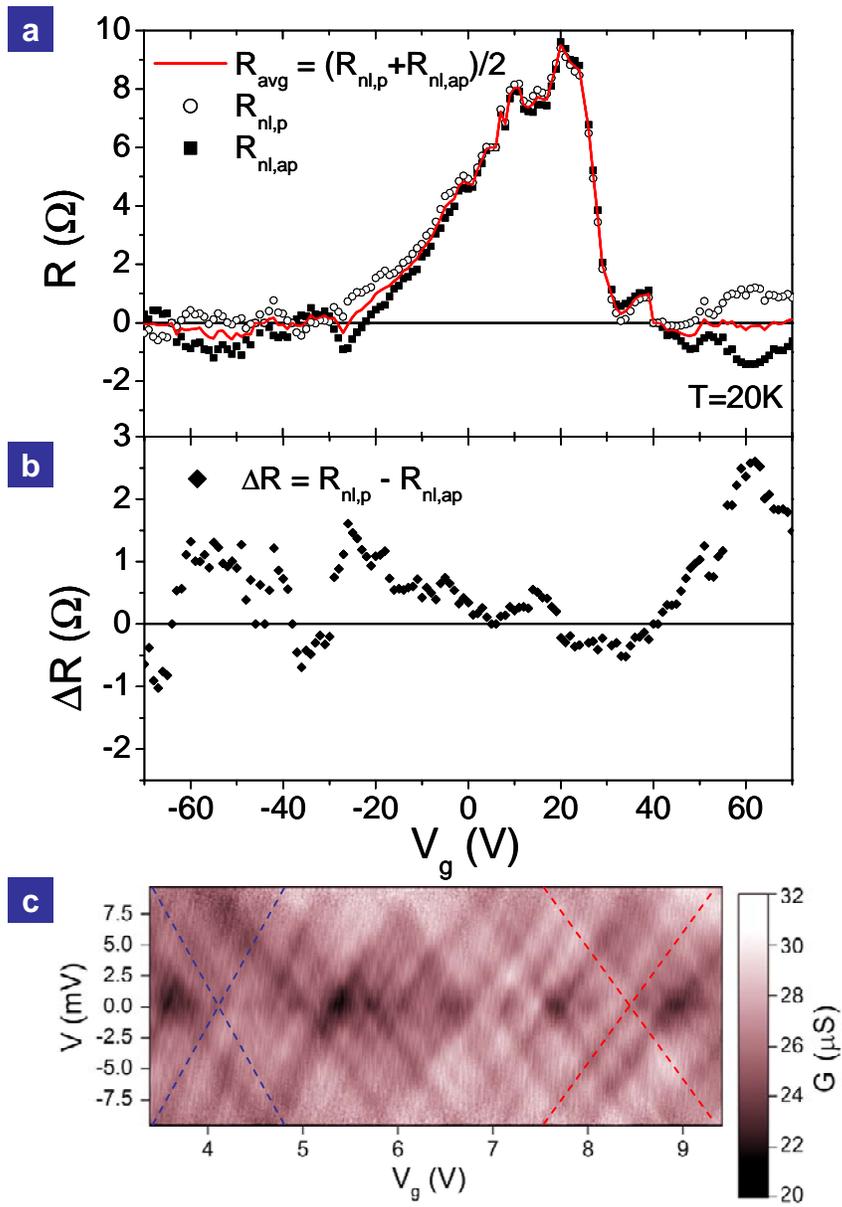

Figure 4

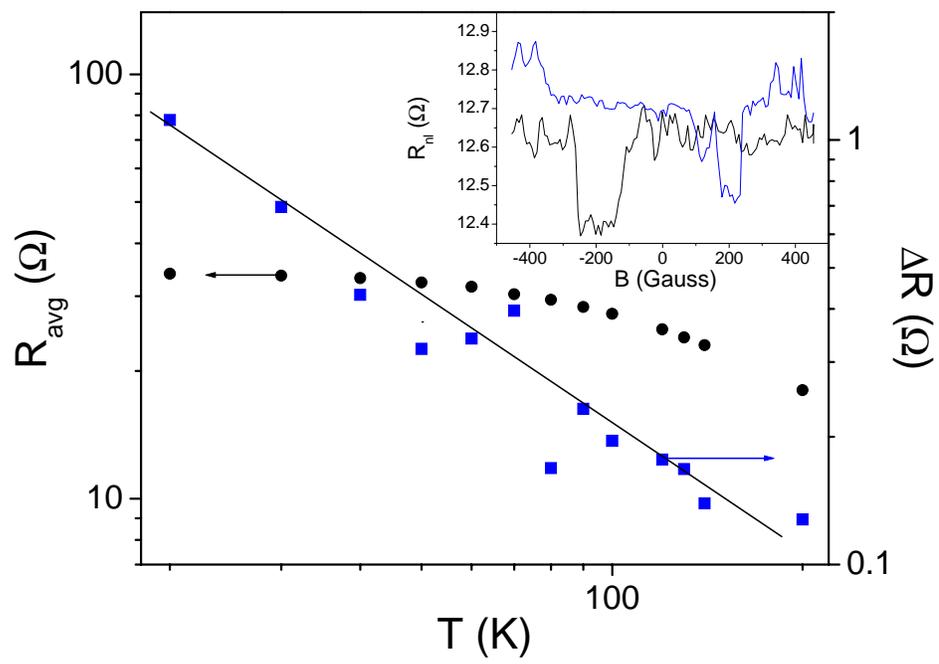